\documentclass[twocolumn,PRA,preprintnumbers,superscriptaddress,amsmath]{revtex4}
\usepackage[latin9]{inputenc}
\usepackage{amsmath}
\usepackage{amssymb}
\usepackage{graphicx}
\usepackage[unicode=true,pdfusetitle,
 bookmarks=true,bookmarksnumbered=false,bookmarksopen=false,
 breaklinks=false,pdfborder={0 0 1},backref=false,colorlinks=false]
 {hyperref}
\hypersetup{
 colorlinks,linkcolor=red,citecolor=blue}

\makeatletter

\DeclareRobustCommand{\greektext}{%
  \fontencoding{LGR}\selectfont\def\encodingdefault{LGR}}
\DeclareRobustCommand{\textgreek}[1]{\leavevmode{\greektext #1}}
\DeclareFontEncoding{LGR}{}{}
\DeclareTextSymbol{\~}{LGR}{126}

\@ifundefined{textcolor}{}
{%
 \definecolor{BLACK}{gray}{0}
 \definecolor{WHITE}{gray}{1}
 \definecolor{RED}{rgb}{1,0,0}
 \definecolor{GREEN}{rgb}{0,1,0}
 \definecolor{BLUE}{rgb}{0,0,1}
 \definecolor{CYAN}{cmyk}{1,0,0,0}
 \definecolor{MAGENTA}{cmyk}{0,1,0,0}
 \definecolor{YELLOW}{cmyk}{0,0,1,0}
}

\makeatother

\begin{document}

\title{Experimental realization of optomechanically induced non-reciprocity }

\author{Z. Shen}

\affiliation{Key Laboratory of Quantum Information, Chinese Academy of Sciences,
University of Science and Technology of China, Hefei 230026, P. R.
China.}

\affiliation{Synergetic Innovation Center of Quantum Information and Quantum Physics,
University of Science and Technology of China, Hefei, Anhui 230026,
P. R. China.}

\author{Y.-L. Zhang}

\affiliation{Key Laboratory of Quantum Information, Chinese Academy of Sciences,
University of Science and Technology of China, Hefei 230026, P. R.
China.}

\affiliation{Synergetic Innovation Center of Quantum Information and Quantum Physics,
University of Science and Technology of China, Hefei, Anhui 230026,
P. R. China.}

\author{Y. Chen}

\affiliation{Key Laboratory of Quantum Information, Chinese Academy of Sciences,
University of Science and Technology of China, Hefei 230026, P. R.
China.}

\affiliation{Synergetic Innovation Center of Quantum Information and Quantum Physics,
University of Science and Technology of China, Hefei, Anhui 230026,
P. R. China.}

\author{C.-L. Zou}

\email{clzou321@ustc.edu.cn}

\affiliation{Key Laboratory of Quantum Information, Chinese Academy of Sciences,
University of Science and Technology of China, Hefei 230026, P. R.
China.}

\affiliation{Synergetic Innovation Center of Quantum Information and Quantum Physics,
University of Science and Technology of China, Hefei, Anhui 230026,
P. R. China.}

\affiliation{Department of Electrical Engineering, Yale University, New Haven,
Connecticut 06511, USA}

\author{Y.-F. Xiao}

\affiliation{State Key Laboratory for Mesoscopic Physics and School of Physics,
Peking University, Beijing 100871, People\textquoteright s Republic
of China.}

\author{X.-B. Zou}

\affiliation{Key Laboratory of Quantum Information, Chinese Academy of Sciences,
University of Science and Technology of China, Hefei 230026, P. R.
China.}

\affiliation{Synergetic Innovation Center of Quantum Information and Quantum Physics,
University of Science and Technology of China, Hefei, Anhui 230026,
P. R. China.}

\author{F.-W. Sun}

\affiliation{Key Laboratory of Quantum Information, Chinese Academy of Sciences,
University of Science and Technology of China, Hefei 230026, P. R.
China.}

\affiliation{Synergetic Innovation Center of Quantum Information and Quantum Physics,
University of Science and Technology of China, Hefei, Anhui 230026,
P. R. China.}

\author{G.-C. Guo}

\affiliation{Key Laboratory of Quantum Information, Chinese Academy of Sciences,
University of Science and Technology of China, Hefei 230026, P. R.
China.}

\affiliation{Synergetic Innovation Center of Quantum Information and Quantum Physics,
University of Science and Technology of China, Hefei, Anhui 230026,
P. R. China.}

\author{C.-H. Dong}

\email{chunhua@ustc.edu.cn}

\affiliation{Key Laboratory of Quantum Information, Chinese Academy of Sciences,
University of Science and Technology of China, Hefei 230026, P. R.
China.}

\affiliation{Synergetic Innovation Center of Quantum Information and Quantum Physics,
University of Science and Technology of China, Hefei, Anhui 230026,
P. R. China.}

\maketitle
\textbf{Non-reciprocal devices, such as circulators and isolators,
are indispensable components in classical and quantum information
processing in an integrated photonic circuit \cite{Shoji2014}. Aside
from those applications, the non-reciprocal phase shift is of fundamental
interest for exploring exotic topological photonics \cite{Lu2014},
such as the realization of chiral edge states and topological protection
\cite{Wang2009,Hafezi2011}. However, incorporating low optical-loss
magnetic materials into a photonic chip is technically challenging
\cite{Bi2011}. In this study, we experimentally demonstrate non-magnetic
non-reciprocity using optomechanical interactions in a whispering-gallery
microresonator, as proposed by Hafezi and Rabl \cite{Hafezi2012}.
Optomechanically induced non-reciprocal transparency and amplification
are observed, and a non-reciprocal phase shift of up to 40 degrees
is demonstrated in this study. The results of this study represent
an important step towards integrated all-optical controllable isolators
and circulators, as well as non-reciprocal phase shifters. }

In recent years, there has been growing interest in the realization
of non-reciprocal photonic devices without magnetic material. To create
genuine non-reciprocal devices, it is necessary to break the Lorentz
reciprocity \cite{Jalas2013}. The methods that attempt to achieve
this goal primarily rely on two mechanisms: the effect due to a moving
medium or reference frame, and the nonlinear optical effect. The Sagnac
effect utilizes the first mechanism and can induce non-reciprocal
phase shift for a photon propagating in a rotating non-inertial frame
of reference \cite{Post1967}. In particular, Fleury et al. \cite{Fleury2014}
reported the sound analogue of isolation due to a circulating fluid
medium. However, the optical Sagnac effect requires a long optical
path, which is not suitable for a photonic chip.
\begin{figure}
\includegraphics[clip,width=7cm]{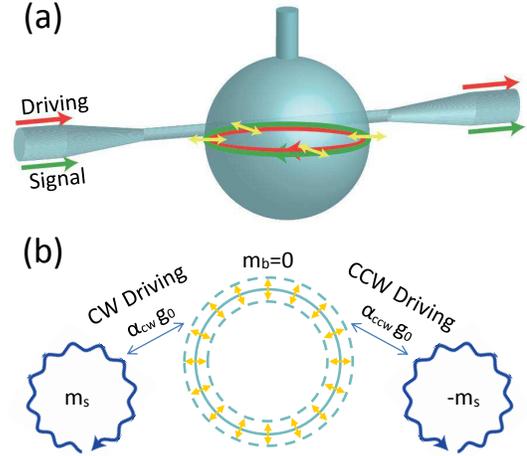}\caption{\textbf{Schematic illustration of optomechanically induced non-reciprocity.}
(a) A strong driving field enhances the optomechanical coupling between
a vibrational mode and the co-propagating optical mode inside a microcavity.
(b) The schematic of clockwise and counter-clockwise optical mode
coupling with the breathing mode. }

\label{Fig1}
\end{figure}

Based on the nonlinear optical effect, the non-reciprocity of a photon
can be achieved by interacting with traveling wave excitations or
equivalent spatiotemporal modulation \cite{Yu2009,Lira2012,Tzuang2014}.
For example, a directional traveling acoustic wave can only efficiently
couple with light that propagates forward or backward due to Brillouin
scattering, which leads to time-reversal symmetry breaking of light
\cite{Kang2011,Dong2015,Kim2015}. A similar effect can also be realized
with a directional photon-photon interaction in a nonlinear micro-ring
resonator \cite{Guo2015}. However, the energy conservation and momentum
matching conditions must be satisfied for all involved modes, resulting
in challenges for dispersion engineering and fabrication of photonic
structures. There are also efforts to create an optical isolator via
Kerr-like nonlinearity induced bistability, but the performance for
a single-photon-level signal and the isolation from weak backward
noise are under debate \cite{Shi2015}.

\begin{figure*}
\includegraphics[width=1.6\columnwidth]{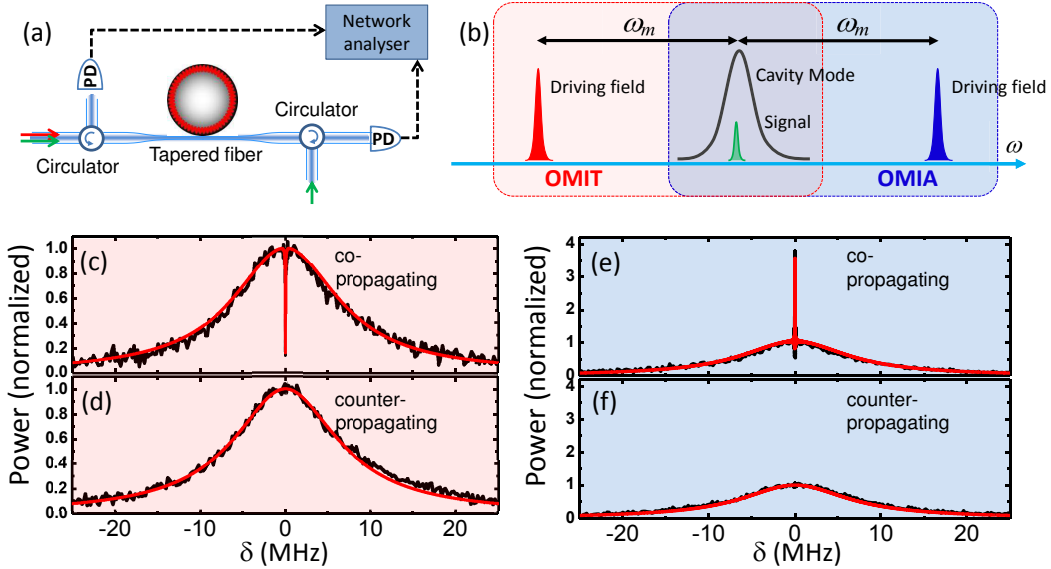}\caption{\textbf{Optomechanically induced transparency (OMIT) and amplification
(OMIA).} (a) A simplified schematic of the experimental setup with
a co- or counter-propagating signal in a silica microsphere. (b) Spectral
position of the driving and signal field in OMIT and OMIA processes.
(c-d) The emission power spectra in the OMIT response are obtained
using a co- and counter-propagating signal pulse. The incident driving
power is 10 mW. The solid red lines are the results of calculations
using the parameters $\omega_{m}/2\pi=88.54$ MHz, $\kappa/2\pi=15$
MHz, $\gamma_{m}/2\pi=22$ kHz, and $C_{\mathrm{cw}}=1.5$, respectively.
(e-f) The emission power spectra in the OMIA response are obtained
using a co- and counter-propagating signal pulse. The incident driving
power is 5.5 mW. The solid red lines are the results of calculations
using the parameters $\omega_{m}/2\pi=88.54$ MHz, $\kappa/2\pi=15$
MHz, $\gamma_{m}/2\pi=22$ kHz, and $C_{\mathrm{cw}}=-0.8$, respectively. }

\label{Fig2}
\end{figure*}

In this article, we report the experimental demonstration of optomechanically
induced non-reciprocity in a silica microsphere resonator. This approach
was first suggested by Hafezi and Rabl \cite{Hafezi2012} and states
that an optical mode dispersively couples with a symmetric radial
breathing mechanical mode. A signal photon can only be affected when
it propagates in the same direction as the driving laser. Compared
to the triple resonant interaction, which consists of two optical
modes and one traveling acoustic wave, in Brillouin scattering \cite{Kang2011,Dong2015,Kim2015},
non-reciprocity can be achieved for any whispering gallery modes in
the cavity considered in this study, which is more favorable for future
applications. Note that this optomechanically induced non-reciprocity
is applicable to all traveling wave resonators and can be easily implemented
on photonic chips. With the mechanical vibrations being cooled to
their ground states, applications in the quantum regime, such as single-photon
isolators and circulators, also become possible.

\begin{figure*}
\centerline{\includegraphics[clip,width=1.6\columnwidth]{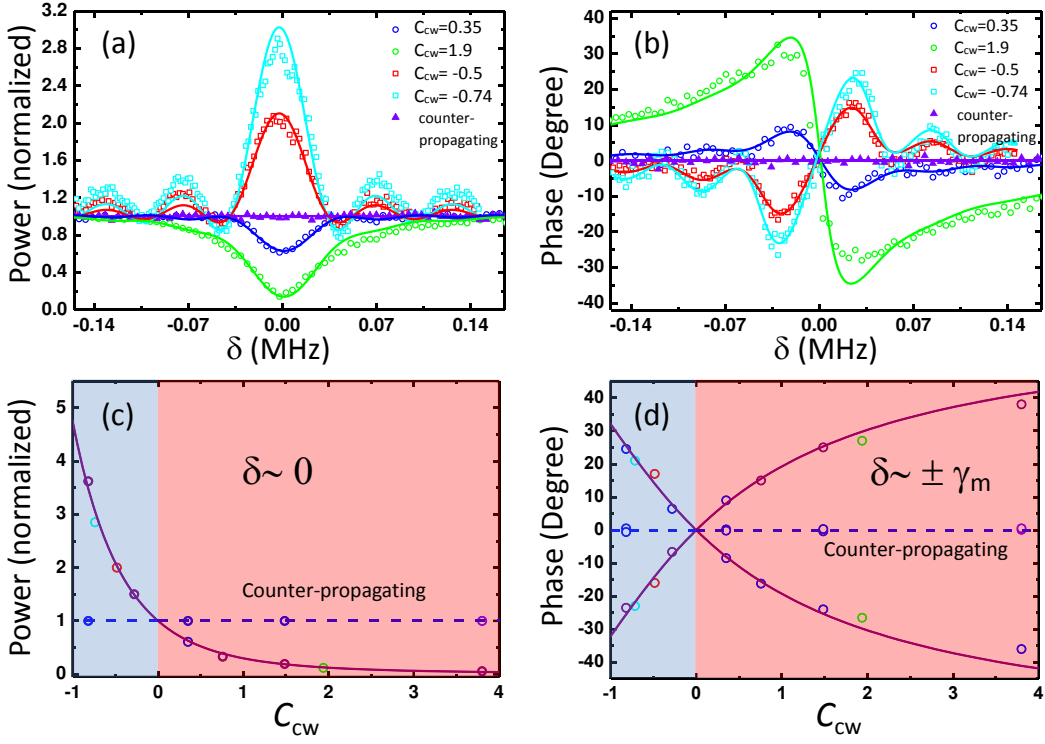}}\caption{\textbf{Optomechanically induced non-reciprocal transmission and phase
shift.} (a) The typical emission power from the microcavity with the
co-propagating (i.e., circle for OMIT and square for OMIA) and counter-propagating
signal (i.e., triangle) and the corresponding phase response (b) at
different driving powers. The solid lines are the results of calculations
using $C_{\mathrm{cw}}=-0.74,-0.5,0,0.35,1.9$.  (c) The spectral
depth of the transparency window or spectral peak of the opacity peak
as a function of the cooperativity $C_{\mathrm{cw}}$. (d) The shift
phase obtained at the detuning $\delta=\pm\gamma_{m}$. The lines
in (c-d) represent the theoretically expected values. }

\label{Fig3}
\end{figure*}

Figure$\,$\ref{Fig1}(a) schematically illustrates a traveling wave
optomechanical system that consists of a whispering gallery microresonator
that is evanescently coupled with a tapered fiber \cite{Park2009,Dong2012,Schliesser2014}.
The traveling wave nature of the microresonator produces a rise in
the degenerate clockwise (CW) and counter-clockwise (CCW) traveling-wave
whispering-galley modes (WGMs), which can also be represented by the
orbital angular momentums $m$ and $-m$, respectively. Similarly,
the mechanical modes in such axial symmetric geometry have an orbital
angular momentum $m_{b}$, where $m_{b}=0$ represents the breath
vibration mode, where the equator of the sphere expands and compresses
uniformly. The variation of the cavity radius leads to the modification
of the resonant frequency of all optical modes \cite{Ma2007,Park2009,Dong2012},
which can be described by the dispersive optomechanical interaction
Hamiltonian \cite{Hafezi2012}:
\begin{equation}
H_{int}=g_{0}(a_{\mathrm{cw}}^{\dagger}a_{\mathrm{cw}}+a_{\mathrm{ccw}}^{\dagger}a_{\mathrm{ccw}})(b+b^{\dagger}),
\end{equation}
where $a_{\mathrm{cw}(\mathrm{ccw})}$ and $b$ denote the Bosonic
operators of the CW(CCW) optical mode and mechanical mode, respectively.
For such an interaction, only a single optical mode (either $a_{\mathrm{cw}}$
or $a_{\mathrm{ccw}}$) is involved; therefore, the properties of
the driving and signal optical fields should be identical, except
for the frequency difference. Thus, the CW(CCW) driving field can
only stimulate the interaction between a phonon and a CW(CCW) signal
photon, as shown in Fig.$\,$\ref{Fig1}(b). This relation between
the driving-signal directions can also be interpreted using the conservation
of orbit angular momentum $m_{s}-m_{d}=m_{b}=0$, where $m_{s}$ and
$m_{d}$ denote the momentum for the signal photon and driving laser,
respectively. As a result, the directional driving field breaks the
time-reversal symmetry and leads to non-reciprocal transmittance for
the signal light: the co-propagating signal photons can be coherently
coupled with phonons, while the coupling between counter-propagating
photons and the phonon is negligible.

The experimental demonstration of the optomechanically induced non-reciprocity
is performed using the experimental setup shown in Fig.$\,$\ref{Fig2}(a).
The driving laser excites the CW optical mode, while either CW or
CCW signal photons are sent to the cavity (see the Supplementary information
for more details with regard to the experimental setup). In a sphere
with a diameter of approximately $36\,\mathrm{\mu m}$, we choose
a high quality factor WGM near 780 nm (linewidth $\kappa/2\pi=15\,\mathrm{MHz}$).
The radial breathing mechanical mode in this sphere has a frequency
of $\omega_{m}/2\pi=88.5\,\mathrm{MHz}$ and a linewidth of $\gamma_{m}/2\pi=22\,\mathrm{kHz}$.

Figure$\,$\ref{Fig2}(b) shows the two configurations of the driving
field. For the driving field that is red-detuned from the cavity
mode by $\omega_{m}$, an effective photon-phonon beam-splitter-like
interaction $(a_{\mathrm{cw}}^{\dagger}b+a_{\mathrm{cw}}b^{\dagger})$
would lead to coherent conversion and induce a transparent window
in the cavity field spectrum \cite{Weis2010,Safavi-Naeini2011}. In
this study, we send a short drive laser and signal pulse (duration
$\tau_{p}=18\,\mathrm{\mu s}$) to measure the transient optomechanical
coupling to avoid thermal instability in the microsphere. The experimental
result in Fig.$\,$\ref{Fig2}(c) shows a sharp transparency window
for co-propagating signal photons, demonstrating the destructive interference
between the signal field and the optical field generated from the
anti-Stokes scattering process. Conversely, the spectrum of the counter-propagating
signal in Fig.$\,$\ref{Fig2}(d) does not show such an effect, which
is indicative of optomechanically induced non-reciprocity. The spectra
still agree with the theoretical predictions of the steady-state intracavity
field\cite{Dong2012}:
\begin{align}
a_{\mathrm{cw}}\left(\delta\right) & =\frac{-\sqrt{\kappa_{in}}\epsilon_{s,\mathrm{cw}}}{i\delta-\frac{\kappa}{2}(1+\frac{C_{\mathrm{cw}}}{1-i2\delta/\gamma_{m}})},\label{eq:spectrum}\\
a_{\mathrm{ccw}}\left(\delta\right) & =\frac{-\sqrt{\kappa_{in}}\epsilon_{s,\mathrm{ccw}}}{i\delta-\frac{\kappa}{2}}.
\end{align}
where $C_{\mathrm{cw}}=\frac{4\left|G_{\mathrm{cw}}\right|^{2}}{\kappa\gamma_{m}}$
is the cooperativity; $G_{\mathrm{cw}}=\sqrt{N_{d}}g_{0}$ is the
effective optomechanical coupling rate; $N_{d}$ is the CW driving
intracavity photon number; $\epsilon_{s,\mathrm{cw}}$ and $\epsilon_{s,\mathrm{ccw}}$
are the weak signal amplitudes of the CW- and CCW- circulating optical
modes, respectively; $\delta$ is the detuning between the signal
and cavity field; and $\kappa_{in}$ is the coupling rate between
the microfiber and microsphere. For a critically coupled optical mode,
the observed intracavity fields for the CW and CCW signal corresponding
to non-reciprocal transmissions of the CW and CCW signal are \textbf{$0$}
and $1$ around $\delta\sim0$, respectively.

For the driving field that is blue-detuned from the cavity mode by
$\omega_{m}$ (i.e., driving frequency $\omega_{d}=\omega_{c}+\omega_{m}$),
an effective photon-phonon pair generation process $(a_{\mathrm{cw}}^{\dagger}b^{\dagger}+a_{\mathrm{cw}}b)$
would lead to optomechanically induced amplification. Figures$\,$\ref{Fig2}(e)
and (f) present the corresponding experimental results, which show
non-reciprocal amplification of the signal. The sequences of driving
laser and signal pulses are the same as before. Here, the experimental
results are fitted by the modified transient intracavity field spectra
because of the unsteady state optomechanical coupling even with 18
\textgreek{m}s driving pulse at blue sideband:
\begin{align}
a_{\mathrm{cw}}\left(\delta\right) & =\frac{-\sqrt{\kappa_{in}}\epsilon_{s,\mathrm{cw}}}{i\delta-\frac{\kappa}{2}(1+\frac{C_{\mathrm{cw}}}{1-i2\delta/\gamma_{m}})}[1+F(\tau_{p})],\label{eq:transient}
\end{align}
with the transient modification $F(t)=\frac{2C_{\mathrm{cw}}Exp\left(i\delta t-(\kappa+\gamma_{m})t/4\right)}{1-i2\delta/\gamma_{m}}\sinh\left[\frac{t\sqrt{\left(\kappa+\gamma_{m}\right)^{2}/4-\kappa\gamma_{m}\left(1+C_{\mathrm{cw}}\right)}}{2}\right]$
. In this study, $C_{\mathrm{cw}}=-\frac{4\left|G_{\mathrm{cw}}\right|^{2}}{\kappa\gamma_{m}}$
has a negative sign, which would effectively narrow the linewidth
of the cavity and amplify the intracavity signal .

To obtain a more quantitative analysis of the optomechanically induced
non-reciprocity, the detailed spectra for different experimental conditions
are shown in Fig.$\,$\ref{Fig3}(a). For blue detuning, the peak
height increases and the linewidth decreases with the driving power.
For red detuning, the transparent depth and linewidth both increase
with the driving power. The intensity at $\delta=0$ and the corresponding
cooperativity $C_{\mathrm{cw}}$ are summarized and plotted in Fig.$\,$\ref{Fig3}(c).
The intensity of the co-propagating signal agrees with the prediction,
while the counter-propagating signal is found to be independent of
$C_{\mathrm{cw}}$.

\begin{figure}
\centerline{\includegraphics[clip,width=8cm]{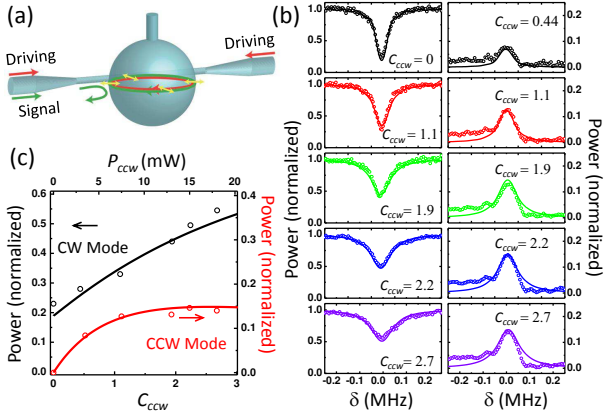}}

\caption{\textbf{Optical mode conversion between two oppositely propagating
optical fields.} (a) Two driving fields are applied simultaneously
to induce the reflection of the signal. (b) The cavity emission power
spectra in the CW direction and the reflected power spectra in the
CCW direction vary with the cooperativity $C_{\mathrm{ccw}}$. The
peak power of the CW driving pulse is 10 mW, corresponding to $C_{\mathrm{cw}}=1.5$.
(c) Emission powers from the CW mode (black circles) and CCW mode
(red circles) at $\delta=0$ as a function of $C_{\mathrm{ccw}}$.
The solid lines in (b-c) are results of the theoretical calculations.}

\label{Fig4}
\end{figure}

The non-reciprocal transmittance of the signal can also be inferred
from the phase because $\arg(a_{\mathrm{cw}}(\delta))$ in Eq.$\,$(\ref{eq:spectrum})
varies with the detuning and $C_{\mathrm{cw}}$. The measured phase
is plotted in Fig.3(b); the difference between the co-propagating
and counter-propagating signal unambiguously shows the non-reciprocal
phase shifter due to the optomechanical interaction. The extracted
phases at $\delta=\pm\gamma_{m}$ that are plotted in Fig.3(d) agree
with the theory. The maximum phase shift achieved in the proposed
experiment is approximately $40$ degrees. The oscillations around
the opacity peak in Fig.$\,$\ref{Fig3}(a) and the corresponding
phase response in Fig.$\,$\ref{Fig3}(b) are observed due to the
transient response of the optomechanical system \cite{Dong2013},
as described by $F(t)$.

Optomechanically induced non-reciprocity is actually controllable
using two oppositely propagating driving fields that excite the CW
and CCW modes simultaneously. As a result of the interesting interplay
between the three coupled modes (Fig.$\,$\ref{Fig1}(b)), there is
an optomechanically dark mode \cite{Dong2012} (i.e., a superposition
of the CW and CCW modes), which enables the conversion of optical
fields from the CW mode to the CCW mode. To test the optical mode
conversion, we fixed the CW driving power at 10 mW and varied the
CCW driving power from 0 to 18 mW, while the input signal coupled
only to the CW mode. As shown in Fig.$\,$\ref{Fig4}(b), the excitation
of the CW mode increases with increasing $C_{\mathrm{ccw}}$, and
the OMIT dip for this mode vanishes and is accompanied by a spectral
broadening of the dip. There is light reflected from the system with
a bandwidth smaller than $100\,\mathrm{kHz}$; its peak power at $\delta\sim0$
also increases with $C_{\mathrm{ccw}}$. The theoretical model predicts
that $a_{\mathrm{cw}}\propto\frac{1+C_{\mathrm{\mathrm{ccw}}}}{1+C_{\mathrm{cw}}+C_{\mathrm{ccw}}}$
and $a_{\mathrm{ccw}}\propto\frac{\sqrt{C_{\mathrm{cw}}C_{\mathrm{ccw}}}}{1+C_{\mathrm{cw}}+C_{\mathrm{ccw}}}$
for the CW signal, which agrees with the experiment results shown
in Fig.$\,$\ref{Fig4}(c). Compared to the CCW signal, for which
$a_{\mathrm{ccw}}\propto\frac{1+C_{\mathrm{cw}}}{1+C_{\mathrm{cw}}+C_{\mathrm{ccw}}}$
and $a_{\mathrm{cw}}\propto\frac{\sqrt{C_{\mathrm{cw}}C_{\mathrm{ccw}}}}{1+C_{\mathrm{cw}}+C_{\mathrm{ccw}}}$,
the co-propagating intracavity excitation and transmission are non-reciprocal
if $C_{\mathrm{CW}}\neq C_{\mathrm{CCW}}$, while the signal reflection
is always reciprocal. Therefore, the system behaves as a controllable
narrowband reflector with non-reciprocal transmittance, which might
be interesting for future studies.

To summarize, optomechanically induced non-reciprocity is experimentally
demonstrated for the first time in this study. The underlying mechanism
of the non-reciprocity demonstrated in this study is actually universal
and can be generalized to any traveling wave resonators via dispersive
coupling with a mechanical resonator, such as the integrated ring-type
microresonator coupled with a nanobeam \cite{Li2009}. Considering
that higher cooperativity and cascading of the non-reciprocal devices
are possible in a photonic integrated chip, optomechanically induced
non-reciprocity has applications in integrated photonic isolators
and circulators \cite{Fu2015}, which will play important roles in
a hybrid quantum Internet \textbf{\cite{Dong2015a}}.

\emph{Note added: When finalizing the manuscript, we became aware
of the related work of F. Ruesink et al. \cite{Ruesink2016}, who
demonstrated optomechanically induced non-reciprocal transmission
in a microtoroid resonator.}

\vbox{}

\noindent

\noindent
\vbox{}

\noindent

\vbox{}

\noindent\textbf{Acknowledgments}\\ The authors would like to thank
H. Wang and X. Guo for discussions. The work was supported by the
Strategic Priority Research Program (B) of the Chinese Academy of
Sciences (Grant No. XDB01030200), the National Natural Science Foundation
of China (Grant No.61308079, 61575184 and 11474011), Anhui Provincial
Natural Science Foundation (Grant No. 1508085QA08), the Fundamental
Research Funds for the Central Universities.

\vbox{}

\noindent

\vbox{}

\noindent

\vbox{}

\end{document}